\documentclass{elsart}
\usepackage{indentfirst}
\usepackage[dvips]{epsfig}
\usepackage{graphicx}

\usepackage{amssymb}
\usepackage{amsmath}

\begin{document}

\begin{frontmatter}

\title{75\%-efficiency blue generation from an intracavity PPKTP frequency doubler}




\author{R. Le Targat, J.-J. Zondy, P. Lemonde}

\address{BNM-SYRTE, Observatoire de Paris\\ 61, avenue de l'observatoire, 75014 Paris, France}

\begin{abstract}
We report on a high-efficiency 461\,nm blue light conversion from
an external cavity-enhanced second-harmonic generation of a
$922\,$nm diode laser with a quasi-phase-matched KTP crystal
(PPKTP). By choosing a long crystal ($L_C=20\,$mm) and twice
looser focusing ($w_0=43\,\mu$m) than the "optimal" one, thermal
lensing effects due to the blue power absorption are minimized
while still maintaining near-optimal conversion efficiency. A
stable blue power of $234\,$mW with a net conversion efficiency of
$\eta=75\%$ at an input mode-matched power of $310\,$mW is
obtained. The intra-cavity measurements of the conversion
efficiency and temperature tuning bandwidth yield an accurate
value $d_{33}(461\,\text{nm})=15\,(\pm 5\%)\,$pm/V for KTP and
provide a stringent validation of some recently published linear
and thermo-optic dispersion data of KTP.
\end{abstract}

\end{frontmatter}

% main text
\section{Introduction}\label{intro}
Continuous-wave (CW) high-power blue light generation is a key
issue for many applications such as laser printing, RGB color
display or for spectroscopy and cooling of atomic species. Due to
the limited power and tunability of gas lasers (Ar$^+$, HeCd) or
newly developed blue diode laser sources in the blue-UV
spectrum~\cite{nakamura}, the usual procedure is to upconvert
near-IR solid-state or semi-conductor diode lasers either internal
to the laser resonator~\cite{white} or in external enhancement
resonators~\cite{kozlovsky,hemmerich}. In the latter scheme, the
efficiency of the upconversion is usually measured in terms of the
ratio $\eta=P_{2\omega}/P_{\omega}^{\text{in}}$ of the generated
second-harmonic (SH) power to the fundamental field (FF) power
which is mode-matched to the resonator. Several nonlinear
materials can upconvert such lasers, using either
temperature-tuned or angular birefringence phase-matching. The
most widely used one is the large nonlinearity
($d_{\text{eff}}\sim 18\,$pm/V) potassium niobate (KNbO$_3$)
crystal~\cite{gunter,zysset,polzik,bode,klappauf}. For
temperature-tuned noncritical phase-matching, the major drawback
of KNbO$_3$ is the occurrence of a phase transition leading to
repoling near $T=185^\circ$C~\cite{gunter,zysset}, which restricts
upconversion to laser wavelengths longer than $\lambda\sim
920\,$nm. For laser cooling of the $^1S_0-^1P_1$ strontium line at
461~nm, which is the target of the present work, the use of a
non-critically phase-matched KNbO$_3$ at $T\sim 150^\circ$C is
hence not recommended. For shorter FF wavelengths, critical
phase-matching at room-temperature is possible but results in a
deleterious beam walk-off ($\rho\sim 1^\circ$) of the blue wave --
leading to elliptical beam shape or even higher-order transverse
patterns~\cite{klappauf} -- combined with a narrow temperature
bandwidth ($\Delta T\sim 0.5^\circ$C)~\cite{bode}. As an
additional drawback, KNbO$_3$ is subject to blue-induced
photo-chromic damage known as BLIIRA (Blue-Induced Infrared
Absorption~\cite{mabuchi}). This nonlinear loss mechanism,
together with the associated thermal lensing, has limited the
highest reported conversion efficiency to $\eta\sim 80\%$,
yielding $500\,$mW of blue power at $473\,$nm from a Nd:YAG laser
power $P_{\omega}^{\text{in}}\sim 800\,$mW~\cite{bode}.

Alternative widely used materials are LiB$_3$O$_5$ (LBO) or
$\beta$-BaB$_2$O$_4$ (BBO)~\cite{beier,woll,li} but the low
nonlinear coefficients of oxoborate crystals ($d_{\text{eff}}\leq
1\,$pm/V) are not suited to the frequency conversion of low power
sources because it requires a tight control of the round-trip
intracavity loss down to $\leq 1\%$. Recently, we employed a
critically phase-matched KTP in a doubly-resonant sum-frequency
generation (SFG) of a Nd:YAG laser and a low-power AlGaAs diode
laser at 813 nm to produce 120 mW of 461 nm light, but the
conversion efficiency ($\eta<30\%$) was limited by the strong
power imbalance of the two pump sources~\cite{ol-courtillot}. To
allow a more efficient cooling of atomic strontium, a new powerful
and more convenient blue source from direct second-harmonic
generation (SHG) of a master-oscillator-power-amplifier
(AlGaAs-MOPA) delivering an output power of $450\,$mW was then
constructed. But unlike in the experiment in Ref.~\cite{klappauf}
that uses a critically phase-matched KNbO$_3$ semi-monolithic
resonator to generate the same wavelength, we made our choice on
periodically-poled potassium titanyl phosphate (PPKTP).
Electric-field poled quasi-phase-matched (QPM) oxide
ferroelectrics such as PPLN (periodically-poled lithium niobate)
and PPKTP have recently super-seeded the previous birefringence
phase-matched materials for visible light
generation~\cite{pierrou,arie,walther} owing to their much higher
effective nonlinearities ($d_{\text{eff}}(\text{PPLN})\sim
17-18\,$pm/V and $d_{\text{eff}}(\text{PPKTP})\sim 7-9\,$pm/V).
Furthermore QPM materials are intrinsically free of walkoff. For
blue generation, PPKTP is preferred to PPLN which exhibits strong
photorefractive damage when used at room-temperature. In
Ref.~\cite{pierrou}, an intra-cavity PPKTP frequency-doubled
Nd:YAG laser yielded an output power of $500\,$mW at $473\,$nm
with an internal efficiency of 5.5\%. Green light power of
$117\,$mW was generated by Juwiler {\em et al} from $208\,$mW of
CW Nd:YAG laser with an efficiency of $56.5\%$ in a
semi-monolithic standing-wave resonator, limited by strong thermal
lensing effects induced by the 532 nm residual
absorption~\cite{arie}. In Ref.~\cite{walther} a MOPA diode laser
($0.5\,$W) similar to ours generated $200\,$mW of blue light at
$461\,$nm in a ring enhancement cavity, a value comparable to that
obtained using a similar semi-conductor laser, at identical
wavelength, in a semi-monolithic standing-wave KNbO$_3$
resonator~\cite{klappauf}.

However due to the lower UV bandgap energy of KTP, linear
absorption becomes an issue at wavelengths shorter than 500 nm. A
detailed absorption measurement of flux-grown or
hydrothermally-grown KTP in the spectral range from the bandgap
wavelength (365 nm) to 600 nm revealed a large fluctuation from
sample to sample~\cite{hansson-ktpabs}. At 473 nm for instance,
values of $\alpha$ ranging from 0.034 to $0.085\,$cm$^{-1}$ were
reported. In a recent closely-related PPKTP-SHG experiment pumped
at 846 nm, a value $\alpha(423\,\,\text{nm})=0.10\,\,$cm$^{-1}$
has been measured~\cite{goudarzi}. Strong thermal lensing
effects~\cite{zondy-ags} arising from the blue absorption was the
main limitation of their power efficiency scaling in genuine CW
operation ($\eta=60\%$, corresponding to 225 mW of 423 nm power
for 375mW of mode-matched Ti:sapphire laser). Higher blue power
(400 mW) could be obtained for the same circulating power of
$P_c=5.5\,$W only in pulsed fringe-scanning mode that allows more
efficient heat dissipation within the PPKTP crystal. Such a severe
limitation in CW operation actually arised from the tight focusing
used ($w_0=17\,\mu$m, corresponding to the theoretical optimum of
the single-pass efficiency for the PPKTP crystal length
$L_C=10\,$mm ~\cite{B&K,zondycomp}) and the associated thermal
lens power which scales as $w_0^{-2}$~\cite{zondy-ags}. In true CW
operation, the thermal focal length (which can be as short as a
few cm, see section~\ref{sec: thermal}) experienced by the
circulating fundamental power impedes efficient mode-and-impedance
matching of the input beam. For a symmetric linear resonator for
instance, thermal lensing has been shown to be responsible for the
clamping of the circulating power to a low critical value
$P_c^{\text{crit}}$ corresponding to the collapse of the secondary
thermally-induced waists~\cite{zondy-ags}.

In the present experiment, we deliberately avoid optimal
single-pass focusing to circumvent these thermal lensing effects.
In section~\ref{sec: analysis} we show that owing to the large
nonlinearity of PPKTP, one doesn't require extremely low
intra-cavity linear losses to maintain the conversion efficiency
constant over a wide range of focusing parameters. The latter is
defined by $L=L_C/z_R$, where $L_C$ is the PPKTP length and
$z_R=k_\omega w_0^2/2$ is the Rayleigh range of the cavity mode.
We find that loose focusing to $w_0=40\,\mu$m in a $20\,$mm long
PPKTP crystal still results in a large single-pass efficiency
$\Gamma=P_{2\omega}/P_c^2\sim 2.3\times 10^{-2}\,$W$^{-1}$ and in
a stable CW operation of the resonant cavity at the maximum
available mode-matched power of $P_\omega^{\text{in}}=310\,$mW
with no evidence of serious lensing effect. Using such a strategy,
blue power scaling to half a Watt should be possible with $\sim
0.7$ W of mode-matched input power.

In section~\ref{sec: setup} we briefly describe the experimental
setup, highlighting some measurement procedures aimed at an
accurate determination of important parameters such as the
mode-matching factor $\kappa$ or the circulating power $P_c$.
Section~\ref{sec: gamma} is devoted to the intracavity measurement
of the conversion efficiency that determines the final enhancement
efficiency, taking profit of the TEM$_{00}$ resonator mode
filtering of the fundamental MOPA laser non-Gaussian beam and
making use of the accurate evaluation of the Gaussian beam SHG
focusing function $h$~\cite{zondycomp}. From these measurements,
we derive a consistent value of the $d_{33}$ nonlinear tensor
element of KTP. The comparison of the recorded temperature tuning
curve with the functional dependence given by two of the most
recently published linear and thermo-optic dispersion relations of
KTP~\cite{fradkin,kato2002} shows a perfect agreement between
theory and experiment, providing thus a stringent validation test
of those dispersion relations for PPKTP in the blue/near-IR
spectrum. Having determined all the relevant parameters, we then
present (section~\ref{sec: resonant}) the resonant enhancement
results which are in good agreement with the theoretical
expectation, with a record efficiency of $\eta=75\%$. The paper
ends with a brief thermal effects analysis (section~\ref{sec:
thermal}) which supports that the experiment is indeed not limited
by thermal effects.

\section{Analysis of singly-resonant SHG efficiency versus focusing}\label{sec: analysis}

We start by investigating the dependence of the power conversion
efficiency on the focusing parameter $L$. At zero cavity detuning,
the internal circulating FF power $P_c$ in a singly-resonant ring
resonator is given by~\cite{kozlovsky,ashkin}
\begin{equation}
    \frac{P_c}{P_\omega^{\text{in}}}=\frac{T_1}{\left[1-\sqrt{(1-T_1)(1-\epsilon)(1-\Gamma
    P_c)}\right]^2}\label{eq: pc}
\end{equation}

where $T_1=1-R_1$ is the transmission factor of the input coupler,
$\epsilon$ is the distributed round-trip passive fractional loss
(excluding $T_1$). $\Gamma$, expressed in W$^{-1}$, is the
depletion due to non linear effects. It can be written as the sum
of two terms :
\begin{equation}
\Gamma=\Gamma_{\text{eff}}+\Gamma_{\text{abs}}
\end{equation}
\begin{itemize}
\item $\Gamma_{\text{eff}}$ is the conversion efficiency,
$P_{2\omega}=\Gamma_{\text{eff}} P_c^2$
\item
$\Gamma_{\text{abs}}$ is the efficiency of the Second Harmonic
absorption process, which cannot be neglected here,
$P_{\text{abs}}=\Gamma_{\text{abs}} P_c^2$
\end{itemize}

The net power conversion efficiency $\eta$ calculated from
(\ref{eq: pc}) obeys the implicit equation~\cite{polzik}
\begin{equation}
    \sqrt{\eta}\left[2-\sqrt{1-T_1}\left(2-\epsilon-\Gamma \sqrt{\frac{\eta P_\omega^{\text{in}}}{\Gamma_{\text{eff}}}}\right)\right]^2
    -4T_1\sqrt{\Gamma_{\text{eff}} P_\omega^{\text{in}}}=0. \label{eq: eta}
\end{equation}
Given $\Gamma$, which depends on the focusing and the crystal
length, and given $\epsilon$ and the maximum available
mode-matched $P_\omega^{\text{in}}$, $\eta$ in Eq.~(\ref{eq: pc})
can be optimized against $T_1$ to yield
\begin{equation}
    T_1^{\text{opt}}=\frac{\epsilon}{2}+\sqrt{\left(\frac{\epsilon}{2}\right)^2+\Gamma
    P_\omega^{\text{in}}}\label{eq: t1opt}
\end{equation}

The conversion efficiency $\Gamma_{\text{eff}}$ can be evaluated
using the undepleted pump SHG theory taking linear absorption into
account~\cite{B&K,zondycomp}. For a waist location at the centre
of the crystal it can be written as \cite{zondycomp,absolute}
\begin{eqnarray}
    \Gamma_{\text{eff}} = \frac{P_{2\omega}}{P_c^2}&=&\frac{2\omega^2
    d_{\text{eff}}^2}{\pi\epsilon_0 c^3 n_\omega^2
    n_{2\omega}}L_Ck_\omega
    \exp(-\alpha_{2\omega}L_C)\,h(a,L,\sigma) \label{eq: gamma}\\
    h(a,L,\sigma)&=&\frac{1}{2L}\int_{-L/2}^{+L/2}\int
    \text{d}\tau\text{d}\tau'
    \frac{\exp[-a(\tau+\tau'+L)-i\sigma(\tau-\tau')]}{(1+i\tau)(1-i\tau')}.
    \label{eq: hfunc}
\end{eqnarray}
In Eqs~(\ref{eq: gamma})-(\ref{eq: hfunc}), $k_\omega=2\pi
n_\omega/\lambda_\omega$ is the FF wavevector internal to the
medium, $\alpha_{n\omega}$ ($n=1,2$) are the linear absorption
coefficients, $a=(\alpha_\omega-\alpha_{2\omega}/2)z_R$,
$L=L_C/z_R$ is the focusing parameter (which differs by a factor 2
from the definition given by Boyd and Kleinman~\cite{B&K}) and
$\sigma=\Delta k\cdot z_R$ is the normalized wavevector mismatch
given by
\begin{equation}
    \Delta k(T)=k_{2\omega}(T)-2k_{\omega}(T)-2\pi/\Lambda(T).
    \label{eq: deltak}
\end{equation}
$\Lambda(T)$ is the PPKTP grating period whose temperature
dependence can be calculated with published thermal expansion
coefficients of KTP~\cite{bierlein,chu-ktp}. We can note that,
when diffraction is considered, the value of $\Delta k$ that
optimizes the focusing function $h$ versus $\sigma$ is not nil as
for a plane wave~\cite{zondycomp}.

The Second Harmonic absorption efficiency $\Gamma_{\text{abs}}$ is
more difficult to express, except in two limits :
\begin{itemize}
\item for a plane wave, one can easily deduce the profile
$P_{2\omega}(z, \alpha_{2\omega})$ for $0 \leq z \leq L_c$ from
Eqs~(\ref{eq: gamma})-(\ref{eq: hfunc}), and thus evaluate the
absorbed power :
\begin{equation}
P_{\text{abs}}=\alpha_{2\omega} \int_0^{L_c} P_{2\omega}(z,
\alpha_{2\omega}) dz \label{eq: Pabs1}
\end{equation}
\item when the beam is tightly focused, the conversion occurs only
at the center of the crystal and then :
\begin{equation}
P_{\text{abs}}=\left(e^{\alpha_{2\omega}\frac{L_c}
{2}}-1\right)\Gamma_{\text{eff}}P_c^2 \label{eq: Pabs2}
\end{equation}
\end{itemize}

With our crystal we measured $\alpha_{2\omega}=0.14\,$cm$^{-1}$
which leads to $\Gamma_{\text{abs}}/  \Gamma_{\text{eff}}= 0.1$
(resp. 0.13) in the plane wave (resp. tight focusing) limit. Given
the small difference between both values and since we are mainly
interested in focusing below optimal, we take for the whole
analysis the plane wave value\footnote{The predictions of the
model will then be slightly too optimistic for large $L$}.

The other parameters of the model are also taken from experimental
measurements (see below). We have $\epsilon=0.02$ lumping the FF
crystal linear loss (absorption + AR coating) and mirror
reflection loss, a crystal length of $L_C=20\,$mm. Our measured
value for the polar tensor element of KTP $d_{33}=15\,$pm/V is
slightly smaller than~\cite{pelz} and yields
$d_{\text{eff}}=9,5\,$pm/V with the refractive indices
$n_\omega\equiv n_Z(922\text{nm})=1.8364$, $n_{2\omega}\equiv
n_Z(461\text{nm})=1.9188$ at $T=30^\circ$C~\cite{kato2002}. The
mode-matched power is $P_\omega^{\text{in}}=310\,$mW.

\begin{center}
\begin{figure}[t]
\begin{center}
\includegraphics[width=12cm]{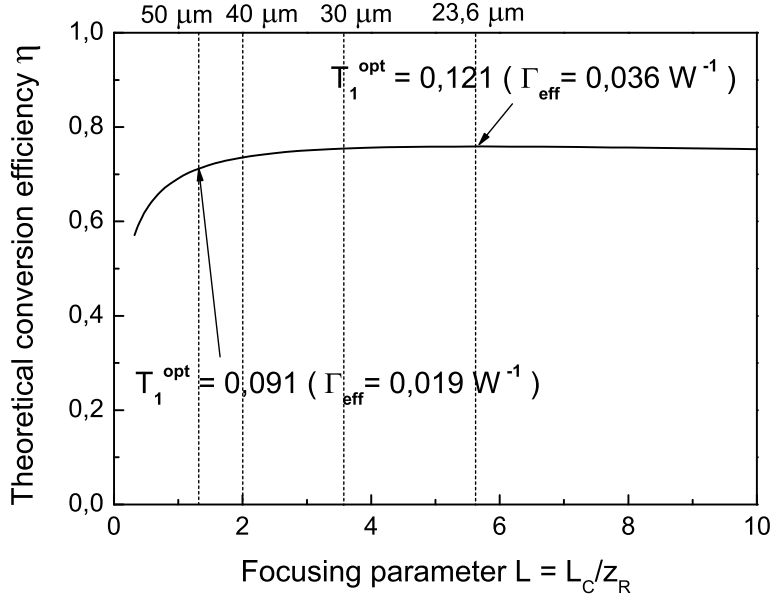}
\end{center}
\caption{\label{fig: eta-vs-focusing} Efficiency factor
$\eta=P_{2\omega}/P_\omega^{\text{in}}$ versus the focusing
parameter when the cavity is impedance-matched
($T_1=T_1^{\text{opt}}$) Parameters are
$P_\omega^{\text{in}}=310\,$mW, $\epsilon=2\%$,
$d_{\text{eff}}=9,5\,$pm/V and $\alpha_{2\omega}=0.14\,$cm$^{-1}$.
Optimal focusing is for $w_0=23.6\,\mu$m, giving a theoretical
conversion efficiency of $76\%$. The labelled arrows indicate the
values of the single-pass conversion efficiencies as calculated
from Eqs.(\ref{eq: gamma})-(\ref{eq: hfunc}).}
\end{figure}
\end{center}

Fig.~\ref{fig: gamma-vs-focusing} displays the efficiency curve
$\eta(T_1^{\text{opt}},L)$ for a beam waist range 18 $\mu$m $\leq
w_0 \leq$ 100 $\mu$m. It corresponds to perfect impedance matching
($T_1=T_1^{\text{opt}}$, see Eq.~(\ref{eq: t1opt})) for each
$\Gamma(L)$. It is clearly seen that $\eta$ is practically
constant over the range $20\,\mu\text{m}\leq w_0\leq 50\,\mu$m,
meaning that it is not necessary to set the cavity waist at the
optimal single-pass conversion as commonly believed, in spite of
the two fold reduction of $\Gamma$ for the loose focusing end
range. At $w_0^{\text{opt}}=23.6\,\mu$m, the circulating optimal
FF power is $P_c=2.56\,$W (yielding $P_{2\omega}=236\,$mW) whereas
at $w_0=50\,\mu$m, $P_c=3.4\,$W ($P_{2\omega}=220\,$mW). For these
nearly identical blue power, the thermal lens power is 4 times
larger at optimal focusing than at the looser focusing. Hence to
avoid thermal effects, a cavity waist between $40\,\mu$m and
$50\,\mu$m would be recommended ($L\leq 2$) with an expected
efficiency $\eta>70\%$.

This insensitivity of $\eta$ as a function of $w_0$ is due to the
large nonlinear efficiency which dominates round trip passive
losses, even for the not so small $\epsilon=0.02$ chosen here.
Experimentally, we indeed found that nearly optimal $\eta$ was
measured over a large cavity waist values, the best trade-off
giving tolerable thermal effects being at $w_0=43\,\mu$m. In the
same way, the value of $T_1$ is not critical. Our input coupler
exhibits $T_1=0.12\,\%$, which causes a loss smaller than 1$\%$ on
the generated blue power in comparison to
$T_1^{\text{opt}}=0.10\,\%$ which would optimize $\eta$ for
$w_0=43\,\mu$m (Eq (\ref{eq: t1opt})).

\section{Experimental setup and measurement procedures}\label{sec: setup}

\begin{center}
\begin{figure}[t]
\begin{center}
\includegraphics[width=12cm]{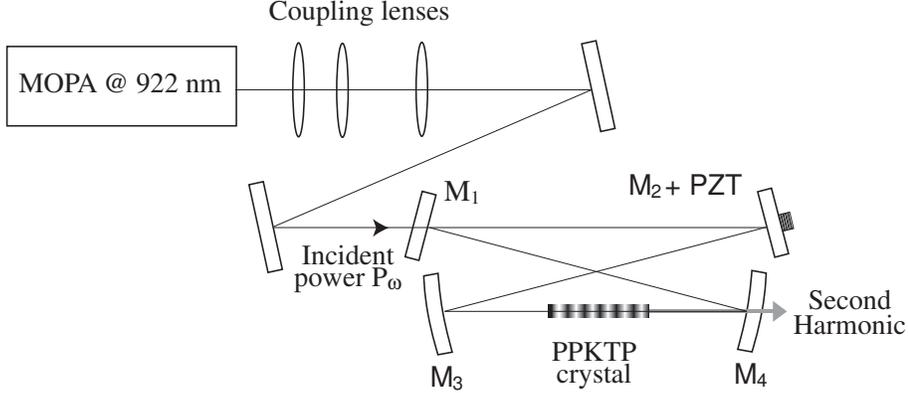}
\end{center}
\caption{\label{fig:schema} Schematic of the experimental setup.}
\end{figure}
\end{center}

The frequency-doubling setup is sketched in Fig.~\ref{fig:schema}.
A conventional unidirectional ring cavity is chosen. The
commercial MOPA pump laser is made of a grating-tuned
extended-cavity master diode laser in the Littrow configuration,
injecting a tapered semiconductor amplifier (Toptica Photonics
AG). After a -70\, dB Faraday isolation stage, the MOPA provides a
useful single-longitudinal-mode power $P_\omega=450\,$mW at
$\lambda_\omega=922\,$nm, with a short-term linewidth of less than
1MHz. The transverse beam shape is far from a fundamental Gaussian
mode and strongly depends on the tapered amplifier injection
current. A system of lenses mode-matches the input FF beam to the
bow-tie ring resonator larger waist located between the 2 plane
mirrors M1 and M2. The folding angle of the ring resonator is set
to $11^\circ$ leading to a negligible astigmatism introduced by
the two off-axis curved mirrors M3 and M4 (of radius-of-curvature
100 mm). The meniscus shape of M4 (M3) do not introduce any
additional divergence of the transmitted FF or SH beam. This
allows an accurate measurement (to $\pm\,5\%$) of the smaller
waist $w_0$ located at the center of the PPKTP crystal from a
z-scan measurement of the TEM$_{00}$ diverging FF beam leaking out
from M4. The diverging beam diameter measurement is performed at
different z-location by use of a rotating knife-edge commercial
device, and the cavity waist is retrieved from standard Gaussian
optics laws.
The pump beam is coupled in through the partial reflector M1.
Mirrors M2-M3 are high-reflectors at 922nm and M4 is
dichroically-coated at $\omega,2\omega$ with $R_\omega>99.9\%$ and
$T_{2\omega}=98\%$. Side 2 of all optics are dual-band AR-coated
with $R\leq 0.5\%$. The dual-band AR-coated PPKTP crystal (Raicol
Crystals Ltd) has dimension $2\times 1\times 20\,$mm$^3$, with the
z-propagation direction corresponding to the X-principal axis and
the 1mm thickness sides oriented along the Z-polar axis. A
first-order (50\%-duty-cycle) QPM periodic grating is patterned
along the X-axis, with a period $\Lambda_0\simeq 5.5\,\mu$m for
temperature quasi-phase-matching around $30^\circ$C. The PPKTP
chip is mounted in a small copper holder attached to a
thermo-electric Peltier element with which we servo the crystal
temperature to better than $\pm\,10\,$mK. The crystal Z-axis is
matched with the direction of the electric-field polarization of
the MOPA laser. The blue absorption coefficient of this crystal
$\alpha_{2\omega}$ at 461 nm was measured from a second blue
source, yielding $\alpha_{2\omega}=0.14\,(\pm\,5\%)$ cm$^{-1}$.
This value is larger than the one
($\alpha_{2\omega}=0.10\,$cm$^{-1}$) measured at 423 nm by
Goudarzi et al~\cite{goudarzi} on a 1cm PPKTP sample from a
different manufacturer, confirming thus the large dispersion of
the values from one sample to another.

Finally, because no suitable $\lambda/2$ plate was available in
front of the cavity for the input power variation at constant MOPA
output power, $P_\omega$ is varied with the PA injection current.
As mentioned previously, such a power variation results in
substantial transverse mode change and in turn in a variation of
the mode-matching factor $\kappa$. The coefficient
$\kappa(P_\omega)$ was hence calibrated in the absence of blue
conversion (by tuning the temperature to a zero conversion regime)
and used to rescale the mode-matched power. For the range
$300<P_\omega<450\,$mW, the mode-matching factor is found
practically constant and equal to $\kappa\sim 0.7$. In the
following section, the conversion efficiency
$\Gamma_{\text{eff}}(w_0)$ (Eq.~(\ref{eq: gamma})) will be
measured internal to the cavity. For the measurement of the
circulating FF power $P_c$, the transmissivity of mirror M4 at 922
nm was accurately calibrated to $1.2\,\,10^{-5}$$\pm\,$10\%. The
FF and the SH powers were calibrated with a thermal powermeter
with an uncertainty below $5\,$\%.

\section{Measurement of PPKTP effective nonlinearity and tuning
curve}\label{sec: gamma}

To model the experimental singly-resonant conversion efficiency,
an accurate knowledge of the experimental
$\Gamma_{\text{eff}}(w_0)$ is needed. We choose not to use the
standard single-pass method for the experimental measurement,
since the poor beam quality of the MOPA output would contradict
the Gaussian pump assumption of Eq.(\ref{eq: hfunc}). By placing
the PPKTP inside the resonator, the mode filtering effect of the
cavity provides instead a pure TEM$_{00}$ pump beam with
accurately known waists from the measurement method outlined in
the previous section. Provided that pump depletion can be
neglected, the intracavity power is considered as constant all
along the crystal, and is defined as the solution of Eq (\ref{eq:
pc}).

Several values of $\Gamma_{\text{eff}}$ were measured for a range
of $P_c$ and 3 waists values $w_0=56,\, 43,\, 36\,\mu$m, to yield
respectively $\Gamma_{\text{eff}}=0.017,\, 0.023,\, 0.028 \,(\pm
10\%)\,$W$^{-1}$. The data points of $P_{2\omega}$ versus $P_c^2$
were excellently fitted by a linear function, meaning that the low
depletion assumption holds for all achievable $P_c$'s, which is
confirmed by an \emph{a posteriori} consistency check of the upper
bound value $\Gamma P_c<0.08\ll 1$. In Fig.~\ref{fig:
gamma-vs-focusing}, the resulting $\Gamma_{\text{eff}}(w_0)$ are
plotted against the focusing parameter $L$ along with the
theoretical curve (Eq.(\ref{eq: gamma})).
\begin{center}
\begin{figure}
\begin{center}
\includegraphics[width=10cm]{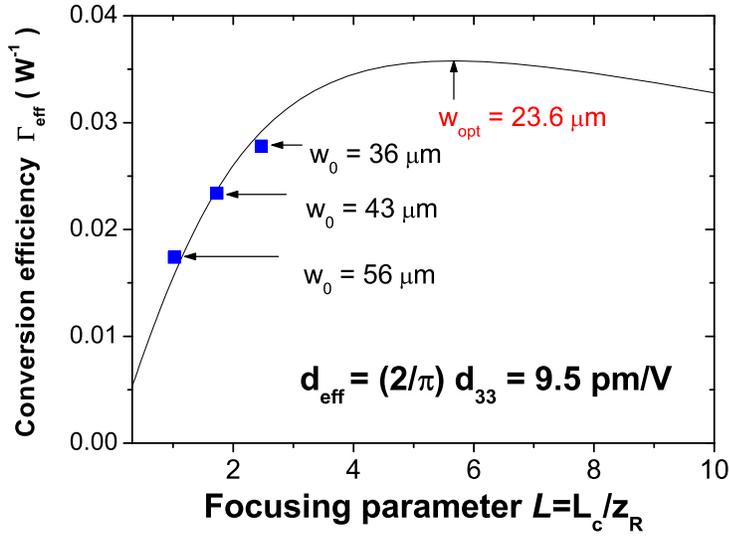}
\end{center}
\caption{\label{fig: gamma-vs-focusing} Conversion efficiencies
$\Gamma_{\text{eff}}$ versus the focusing parameter $L=L_c/z_R$.
The 3 squares are the experimental points measured at
$w_0=56,43,36\,\mu$m and the solid line are the theoretical values
from Eqs.(\ref{eq: gamma})-(\ref{eq: hfunc}) using the labelled
value of $d_{\text{eff}}$ after optimisation of the focusing
function $h$ over the mismatch parameter $\sigma$. The theoretical
optimal waist is found at $w_{\text{opt}}=23.6\,\mu$m.}
\end{figure}
\end{center}
The only adjusted parameter to match the experimental points to
the curve is the effective nonlinear coefficient
$d_{\text{eff}}=(2/\pi)d_{33}$ for a first-order QPM, which is
found to be $d_{\text{eff}}=9.5\,(\pm\,5\%)\,$pm/V. Such a value
is slightly higher than those measured elsewhere (5-8 pm/V) from
OPO threshold measurement or difference frequency
generation~\cite{fradkin,pelz}, even when wavelength dispersion is
accounted for. This value yields for KTP $d_{33}(461nm)=15\,(\pm
\,5\%)$\,pm/V, which is the commonly reported value of this polar
$\chi^{(2)}$ tensor element~\cite{pelz,boulanger,shoji} and
matches exactly the value (14.8 pm/V) reported by the
manufacturer~\cite{eger}. This result shows that the grating
quality of our cristal is extremely high. We believe that the
perfect match of $d_{\text{eff}}$ with its maximum theoretical
value stems from the pure Gaussian beam measurement and analysis
taking diffraction and absorption effects into account, which is
not always the case with some of the reported lower values even
accounting for grating periodicity defects.

We have tried tight focusing close to the optimal waist shown in
Fig.~\ref{fig: gamma-vs-focusing} without any improvement in the
singly-resonant conversion efficiency, despite the larger
$\Gamma_{\text{eff}}$, confirming the prediction of
Section~\ref{sec: analysis}. At $w_0=36\,\mu$m, the conversion
efficiency is identical to the one at $43\,\mu$m. As expected,
increasing thermal effects occurred with smaller waists, which
could be assessed from a broadened triangular bistable shape of
the FF and blue fringes as the cavity length is swept on the
contracting length side of the voltage ramp (see e.g. Fig.~10 of
Ref.~\cite{goudarzi} and the thermal effects analysis in
Section~\ref{sec: thermal}). On this side of the fringe, passive
self-stabilization of the optical path length of the resonator
tends to maintain the cavity in resonance with the incoming FF
frequency. Such an opto-thermal dynamics of thermally-loaded
resonators has been analyzed in detail in Ref.~\cite{zondy-ags}.
Furthermore, the active locking of the cavity to the top of the
distorted fringe - a bistable operating point~\cite{zondy-ags} -
becomes problematic as reported in Ref.~\cite{goudarzi}.
\begin{center}
\begin{figure}
\begin{center}
\includegraphics[width=10cm]{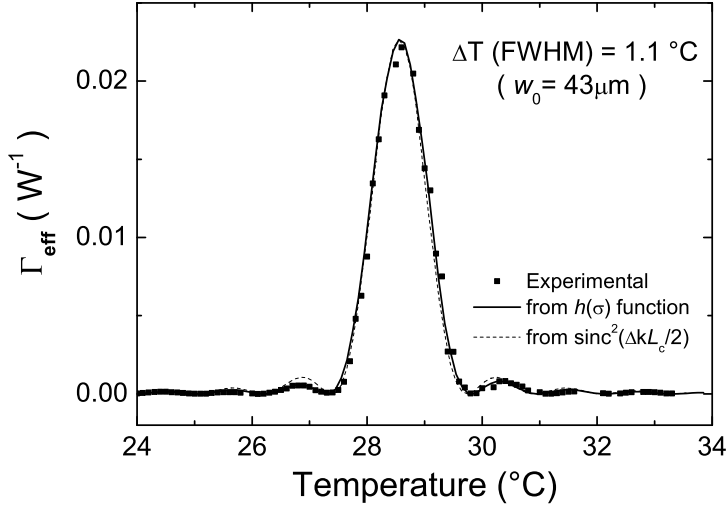}
\end{center}
\caption{\label{fig: temp-tuning} Measured (black circles) and
calculated (solid line) temperature tuning curve at
$w_0=43\,\mu$m. The dashed line corresponds to a plane wave.}
\end{figure}
\end{center}
The temperature tuning curve measured internal to the cavity is
also shown in Fig.~\ref{fig: temp-tuning} for the final waist
choice of $w_0=43\,\mu$m, corresponding to $L=1.73$. The
oscillatory fine structure pattern at the wings is clearly seen.
The solid line is the conversion efficiency computed from
Eqs.(\ref{eq: gamma})-(\ref{eq: deltak}), where the wavevector
mismatches $\Delta k(T)$ have been computed with the Sellmeier
relations at room-temperature given in Ref.~\cite{fradkin}, the
thermo-optic dispersion relation of Ref.~\cite{kato2002} and the
thermal expansion coefficient $\alpha_X=6.8\times 10^{-6}/^\circ$C
reported in Ref.~\cite{chu-ktp}. Given the short grating period, a
strikingly excellent agreement is seen with the data points
(sidelobe amplitudes and positions) when the $h(\sigma)$ focusing
function is used instead of the plane-wave formula. Apart from a
lateral temperature shift of the calculated tuning curve in
Fig.~\ref{fig: temp-tuning}, we stress again that no fit was made
to get such an agreement, which not only provides a stringent
validation of the used dispersion data, at least for PPKTP
produced by our manufacturer, but also highlights the excellent
quality of the first-order periodic grating over the whole PPKTP
length. The FWHM temperature tuning bandwidth is $\Delta
T=1.1^\circ$C.

\section{Resonant enhancement efficiency}\label{sec: resonant}
The final cavity dimensions yielding the least thermal effects
while providing the best power conversion efficiency when the
cavity length is servo-controlled correspond to cavity waists
$(w_0,w_1)=(43\,\mu$m,$163\,\mu$m) given by a M3-M4 spacing of
$\sim 130\,$mm and a total ring cavity round-trip length
$L_{\text{cav}}=569\,$mm. For genuine CW stable operation, an
active electronic servo based on an FM-to-AM fringe modulation
technique was preferred to the optically phase-sensitive
H{\"a}nsch-Couillaud method that was reported to fail when thermal
effects arise~\cite{bode,goudarzi}.

The round-trip intracavity fractional loss $\epsilon$ is measured
by fitting, at the QPM temperature, the total losses
$p=\epsilon\,+\,\Gamma\,P_c$ of the cavity as a function of $P_c$.
Otherwise, $p$ is related to the contrast $C$ of the cavity
reflection fringes :
\begin{equation}
p = \frac {C\,P_{\omega}} {P_c}
\end{equation}
This relationship is very reliable since it doesn't depend on the
mode matching coefficient $\kappa$. The fit yields
$\epsilon=0.021\,(\pm\,5\%)$, which is further checked from the
value of the cavity finesse $\mathcal{F}\simeq 40$ in the absence
of nonlinear conversion. This also gives a measurement of $\Gamma$
which is consistent with the value quoted above.

When the cavity is close to impedance matching, the reflection
contrast becomes nearly constant and is then equal to the mode
matching coefficient, we measured
$\kappa\left(P_{\omega}^{max}\right) = 73\,\%$. For intermediate
values of the pump power, the evaluation is more difficult :
$\kappa$ can be deduced from the contrast of reflected fringes at
zero conversion, once $\epsilon$ is known.

\begin{center}
\begin{figure}
\begin{center}
\includegraphics[width=9cm]{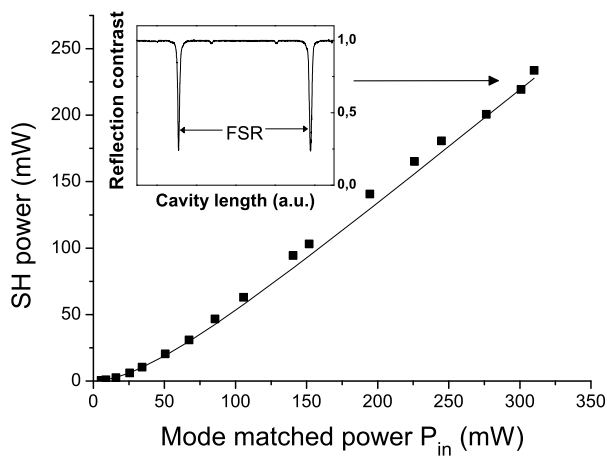}
\end{center}
\caption{\label{fig: p2-vs-pin} SH power versus the mode-matched
FF power. The solid line is a calculation from Eq.~(\ref{eq: eta})
making use of $\epsilon=0.021$, $T_1=0.12$,
$\Gamma=0.023\,$W$^{-1}$. The inset shows the FF reflection fringe
contrast at maximum power.}
\end{figure}
\end{center}
\begin{center}
\begin{figure}
\begin{center}
\includegraphics[width=9cm]{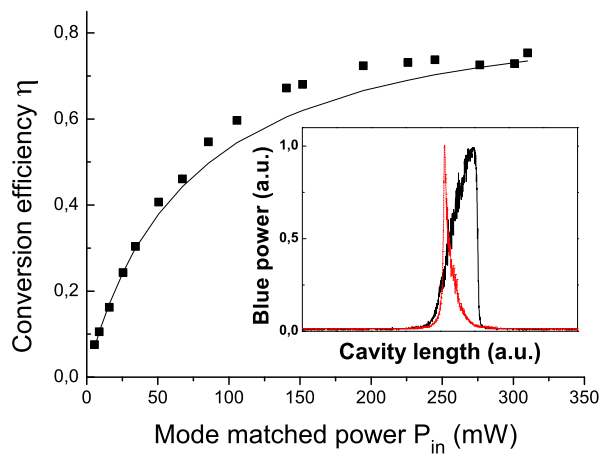}
\end{center}
\caption{\label{fig: eta-vs-pin} SH power efficiency versus the
mode-matched FF power. The solid line is again computed from
Eq.~(\ref{eq: eta}) with the same parameter as for Fig.~\ref{fig:
p2-vs-pin}. The inset shows a magnification of the blue fringe
recorded at maximum power (solid line: contracting cavity length,
dotted line: expanding length). The small difference in their
relative width witnesses low thermal effects as compared with
Ref.~\cite{goudarzi}.}
\end{figure}
\end{center}

The generated blue power and net power efficiency $\eta$ are
plotted against the mode-matched power in Fig.~\ref{fig:
p2-vs-pin} and Fig.~\ref{fig: eta-vs-pin}. The solid lines are
computed from Eq.~(\ref{eq: eta}) using the experimentally
measured parameters. Experimental dots match well this curve,
meaning that thermal effects are not a problem up to the maximum
available laser power. The small offset between experimental
points and the theoretical curve from 100 mW to 250 mW being
probably due to the less accurate evaluation of the mode matched
power in this range. At maximum power, the finesse drops to
$\mathcal{F}\simeq 30$ due to the nonlinear loss (see inset of
Fig.~\ref{fig: p2-vs-pin} which shows the reflected FF fringe with
a contrast of $\sim73\,\%$). The conversion efficiency is
independent on whether the measurement is made under pulsed
scanning-mode or under cavity-locked operation. In the latter case
however, a slight adjustment of the PPKTP temperature has to be
performed to cancel the temperature-induced phase-mismatch under
CW operation (Section~\ref{sec: thermal}). At the maximum
mode-matched input power $P_\omega^{\text{in}}=310\,$mW
($P_{c}=3.2\,$W is the corresponding circulating power), one
obtains $P_{2\omega}=234\,$mW corresponding to $\eta=75\%$.
\\

\section{Blue-induced thermal effects analysis}\label{sec: thermal}
The inset of Fig.~\ref{fig: eta-vs-pin} shows a slight fringe
asymmetry observed on both the FF and SH fringes, reminiscent of
the onset of thermal bistability. On the contracting cavity length
scan (solid line), the fringe is broadened because the
opto-thermal dynamics on this side is characterized by a
self-stabilizing effect of the optical path to the laser frequency
whereas on the expanding length side (dotted curve) the feedback
is positive~\cite{zondy-ags}. In the PPKTP experiment of
Ref.~\cite{goudarzi}, the thermal effects were so prominent that
the fringe shape broadens over several cold cavity linewidths to
acquire a triangular shape. In our case, the broadening is
comparatively very modest. It is expected that heating is due to
both the residual FF absorption and the SH absorption.
%is comparable to what was observed
%in a blue generation SHG experiment with KNbO$_3$~\cite{ludlow}.
To quantify further the role of residual thermal effects, we use
the radial heat diffusion model detailed in Ref.~\cite{zondy-ags},
assuming an equivalent crystal rod radius equal to the
half-thickness of the PPKTP ($r_0=0.5\,$mm). The temperature rise
due to the FF absorption can then be written as $\Delta T\equiv
T-T_{0}=\Delta T_0-\frac{1}{2}\rho r^2$ where $T_0$ is the nominal
phase-matching temperature in the absence of heating and $r$ is
the radial coordinate. The uniform temperature shift $\Delta T_0$
expresses as
\begin{equation}
\Delta T_0=\frac{\alpha_{\omega}P_{c}}{4\pi
K_c}\left[0.57+\ln\left(\frac{2r_0^2}{w_0^2}\right)\right] \equiv
kP_{c}.\label{eq: dt0}
\end{equation}
The quantity $\alpha_\omega P_c$ denotes the absorbed power per
unit length. The quadratic term coefficient in $\Delta T$,
responsible for lensing, is $\rho=\alpha_{\omega}P_{c}/(\pi K_c
w_0^2)$ with a corresponding thermal lens power
\begin{equation}
    p=\frac{1}{f_{\text{th}}}=\frac{P_{\omega}^{\text{abs}}}{\pi
    w_0^2}\left(\frac{\text{d}n_{\omega}/\text{d}T}{K_C}\right)\label{eq: lens}
\end{equation}
where $P_{\omega}^{\text{abs}}=\alpha_{\omega}L_C P_{c}$ is the
total FF absorbed power ($P_{\omega}^{\text{abs}}=19.5\,$mW at
$P_{c}=3.25\,$W for $\alpha_\omega=0.3\,\%\,$cm$^{-1}$). The
quantity in parenthesis,
$\zeta=\frac{\text{d}n_{\omega}/\text{d}T}{K_C}$, is the thermal
figure-of-merit of KTP with
$K_c=3.3\,$W/(m$\cdot^\circ$C)~\cite{bierlein} and
$\text{d}n_{\omega}/\text{d}T=1.53\times 10^{-5}$~\cite{kato2002}.
The normalized FF fringe lineshape under adiabatic length scan
$y(\delta)=P_c(\delta)/P_{cm}$, where $P_{c}(\delta=0)$ is the
maximum intra-cavity power given by Eq.(\ref{eq: pc}), is
described by
\begin{equation}
    y(\delta)=\frac{1}{1+\left[\delta-\Delta y(\delta)\right]^2}.\label{eq: fringe-transmission}
\end{equation}
where $\delta=(\nu_L-\nu_{\text{cav}})/\gamma$ is the normalized
cavity detuning, $\gamma$ being the cold fringe half-linewidth :
$\gamma=c/[L_{\text{cav}}+L_c(n_\omega-1)]/2\mathcal{F}=8.5\,$MHz.

In Eq.~(\ref{eq: fringe-transmission}), the Airy function has been
approximated by a Lorentzian in the vicinity of the resonance. For
$\Delta\neq 0$, the cavity resonance is seen to move adiabatically
as the crystal is thermally loaded. The FWFM of the thermally
broadened resonance is given in unit of $\gamma$ by
$\Delta=\alpha_\omega\mathcal{F}L_C
P_{c}\zeta/2\pi\lambda_\omega$~\cite{zondy-ags}. For a
sufficiently strong thermal load ($\Delta\gg 1$) the fringe shape
presents an hysteresis behavior, the saddle-node point
corresponding to the top of the fringe.

For $P_{c}=3.25\,$W, one finds $\Delta T_0=0.14^\circ$C, and
$f_{\text{th}}=65\,$mm. Such a short focal length means that
significant FF lensing still occurs despite the looser focusing
used and the small value of FF absorption coefficient. Only a
rigorous hot ring cavity waist analysis, similar to the one
developed for a standing-wave symmetric resonator in
Ref.~\cite{zondy-ags}, can give an indication on the influence of
this thermal lens on the cold cavity waist. For a symmetric
resonator, the length must be reduced to the minimum allowed space
in order to contain the thermal lens effect. Due to $\Delta T_0$,
the FF hot fringe is shifted from the cold fringe position by
$\Delta\simeq 0.47$ half-linewidth only, given $\mathcal{F}\sim
30$.

To evaluate the heating due to the SH absorbed power,
Eq.~(\ref{eq: dt0})-(\ref{eq: lens}) cannot be used as they are,
because the blue power is not uniform along the crystal
($P_{2\omega}=0$ at $z=0$ and is maximum at $z=L_c$). It is
however possible to estimate the total blue absorbed power
$P_{2\omega}^{\text{abs}}=\Gamma_{\text{abs}}P_c^2$ from
Eq.~(\ref{eq: Pabs1}), which yields
$P_{2\omega}^{\text{abs}}=25\,$mW, and
$\Gamma_{\text{abs}}=2.4\,10^{-3}\,W^{-1}$. This heating power can
be distributed uniformly if we define an effective absorption
coefficient and an effective incoming power such that
$\alpha_{2\omega}P_{2\omega}=P_{2\omega}^{\text{abs}}/L_c=0.012\,$W/cm.
Making now use of Eq.~(\ref{eq: dt0}) with $w=w_0/\sqrt{2}$, one
is lead to $\Delta T_0=0.34^\circ$C. With both FF and SH heating
sources, the scanning fringe lineshape becomes
\begin{equation}
    y(\delta)=\frac{1}{1+\left[\delta-\Delta y(\delta)-\Theta y^2(\delta)\right]^2}.\label{eq: new-fringe}
\end{equation}
where $\Theta=\Gamma_{\text{abs}}\mathcal{F}
P_{c}^2\zeta/2\pi\lambda_\omega\simeq 0.67$. The above fringe
analysis holds as long as BLIIRA or additional photochromic
phenomena, which induce further FF absorption, is neglected.

\begin{center}
\begin{figure}
\begin{center}
\includegraphics[width=9cm]{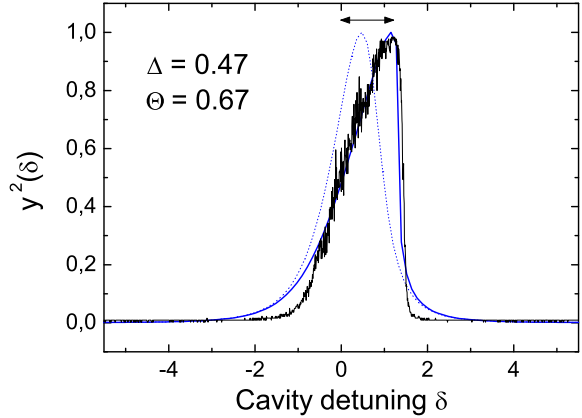}
\end{center}
\caption{\label{fig: blue-fringe-shape} Blue fringe shape under
adiabatic cavity length scan on the negative feedback side of the
triangular voltage ramp. Dotted line: FF thermal load only; solid
line: FF and SH thermal load. The arrow shows the shift $\sim
\Delta+\Theta$ from the cold Lorentzian centered at $\delta=0$
(not shown).}
\end{figure}
\end{center}

The nonlinear equation (\ref{eq: new-fringe}) is solved for $y$ as
a function of the detuning $\delta$ to model the SH fringe shape
$y^2(\delta)$. Figure~\ref{fig: blue-fringe-shape} shows the
experimental blue fringe recorded for a low-rate
contracting-length scan ($f=20\,$Hz), superimposed with the
calculation from Eq~(\ref{eq: new-fringe}) with either $\Theta=0$
(i.e. considering only FF heating, dotted curve) or with both FF
and SH heating (solid line).
From the analysis, the weak thermal dynamics is mainly ruled by
the SH absorption. No bistability is observed meaning that these
thermal effects do not affect the measured conversion efficiency,
contrary to the case of Ref.~\cite{goudarzi} where the fringe
shift amounts to $>15 \gamma$, giving a broad triangular shape and
a sharp spike on the expanding-length scan. Indeed, considering a
shorter $L_c=1\,$cm PPKTP at near optimal focusing ($w_0\sim
20\,\mu$m), a lower value $\Gamma_{\text{eff}}=0.008\,$W$^{-1}$ as
measured in Ref.~\cite{goudarzi} for
$P_\omega^{\text{in}}=360\,$mW, one is lead to the same efficiency
$\eta\sim 73\%$ but with a larger $P_{cm}=5.72\,$W and
$\mathcal{F}=50$. With only twice the intracavity FF power,
$\Delta T_0(\text{FF})=0.32^\circ$C, $f_{\text{th}}=15.7\,$mm and
$\Delta T_0(\text{SH})=2.6^\circ$C.
\begin{center}
\begin{figure}
\begin{center}
\includegraphics[width=9cm]{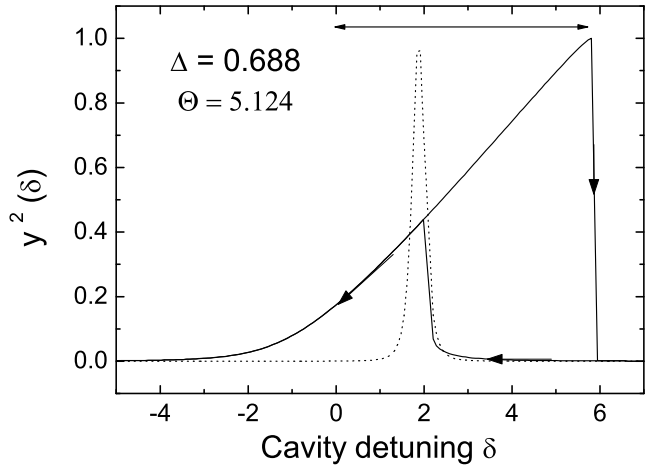}
\end{center}
\caption{\label{fig: strong-fringe-shape} Same calculation as in
Fig.~\ref{fig: blue-fringe-shape} performed here for stronger
thermal effects corresponding to $P_{cm}=5.7\,$W, with
$P_{\omega}^{\text{abs}}=17\,$mW and
$P_{2\omega}^{\text{abs}}=128\,$mW ($L_c=10\,$mm, $w_0=
20\,\mu$m). The dotted line would be the sharp fringe observed on
the reverse scan direction.}
\end{figure}
\end{center}
For this tighter focusing case, the corresponding blue fringe
shape plotted in Fig.~\ref{fig: strong-fringe-shape} displays a
bistable shape. On the contracting-length direction (to the right)
a fringe broadening by $\Delta+\Theta=5.8$ (in unit of $\gamma$)
is evidenced. On the expanding-length direction (to the left), the
opto-thermal feedback is positive. Once the upper intensity branch
is reached, instead of following it back adiabatically as yielded
by the stationary calculation, a sharp fringe results as sketched
by the dotted line.

\section{Conclusion}
We have generated more than 230~mW of blue power at 461~nm
starting from $P_\omega^{\text{in}}=310$~mW of mode-matched
fundamental power from a MOPA diode laser, with a net power
efficiency $\eta=75\%$ using an external-cavity enhanced PPKTP
frequency doubler. The key ingredient for such a high efficiency
is the use of a long crystal length allowing to relax the focusing
parameter to a level where thermal effects due to the strong blue
absorption (temperature mismatch and lensing effects) don't
prevent from a CW operation at maximum power. No cavity locking
issue for stable CW were encountered at maximum laser power. The
long nonlinear interaction length results in a large single-pass
conversion efficiency that relaxes the constraint of having
vanishing round-trip linear loss.

Finally, by measuring the effective nonlinear coefficient internal
to the ring cavity, we have additionally provided an accurate
determination of the $d_{33}(461\,\text{nm})=15\pm (5\%)\,$pm/V
nonlinear tensor element of KTP from exact Gaussian beam SHG
theory accounting for absorption by the crystal. The high-quality
of the first-order QPM grating of our sample was also checked from
the perfect agreement of the measured phase-matching temperature
bandwidth with a set of published Sellmeier equations and
thermo-optic dispersion relations of KTP.

% \appendix

% \section{}
% \label{}


\begin{thebibliography}{00}


% notes:

% \bibitem{label2}
% If there is a note, it should come last:
% \bibitem{label3} \note
% \end{subbibitems}
\bibitem{nakamura} S. Nakamura, G. Fasol, \emph{The Blue Laser
Diodes}, Springer Verlag, Heidelberg, 1997).
\bibitem{white} R.T. White, I.T. McKinnie, S.D. Butterworth, G.W.
Baxter, D.M. Warrington, P.G.R. Smith, G.W. Ross, D.C. Hanna,
Appl. Phys. B~{\bf 77} (2003) 547.
\bibitem{kozlovsky} W.J. Kozlovsky, C.D. Nabors, R.L. Byer, J.
Quantum Electron.~{\bf 24} (1988) 913.
\bibitem{hemmerich} A. Hemmerich, D.H. McIntire, C. Zimmermann,
T.W. H\"ansch, Opt. Lett.~{\bf 15} (1990) 372.
\bibitem{gunter} B. Zysset, I. Biaggio, P. G\"unter, J. Opt. Soc.
Am. B~{\bf 9} (1992) 380.
\bibitem{zysset} I. Baggio, O. Kerkoc, L.-S. Wu, P. G\"unter, B.
Zysset, J. Opt. Soc. Am. B~{\bf 9} (1992) 507.
\bibitem{polzik} E.S. Polzik, H.J. Kimble, Opt. Lett.~{\bf 16}
(1991) 1400.
\bibitem{bode} M. Bode, I. Freitag, A. T\"unnermann, H. Welling,
Opt. Lett.~{\bf 22} (1997) 1220.
\bibitem{klappauf} B.G. Klappauf, Y. Bidel, D. Wikowsky, T.
Chaneli{\`e}re, R. Kaiser, \emph{Detailed study of efficient blue
laser source by second harmonic generation in a semimonolithic
cavity for the cooling of strontium atoms}, Appl. Opt.~{\bf
43}(2004) 2510.

\bibitem{mabuchi} H. Mabuchi, E.S. Polzik, H.J. Kimble, J. Opt.
Soc. Am. B~{\bf 11} (1994) 2023.

\bibitem{beier} B. Beier, D. Woll, M. Scheidt, K.-J. Boller, R.
Wallenstein, Appl. Phys. Lett.~{\bf 71} (1997) 315.
\bibitem{woll} D. Woll, B. Beier, K.-J. Boller, R. Wallenstein, M.
Hagberg, S. O'Brien, Opt. Lett.~{\bf 24} (1999) 691.
\bibitem{li} P. Li, D. Li, Z. Zhang, S. Zhang, Opt. Commun.~{\bf
215}(2003) 159.
\bibitem{ol-courtillot} I. Courtillot, A. Quessada, R.P. Kovacich,
J.-J. Zondy, A. Landragin, A. Clairon, P. Lemonde, Opt. Lett.~{\bf
28} (2003) 468.
\bibitem{pierrou} M. Pierrou, F. Laurell, H. Karlsson, T. Kellner,
C. Czeranovsky, H. Huber, Opt. Lett.~{\bf 24} (1999).
\bibitem{arie} I. Juwiler, A. Arie, Appl. Opt.~{\bf 42} (2003)
7163.
\bibitem{walther} Ch. Schwedes, E. Peik, J. Von Zanthier, A.Y. Nevsky, H.
Walter, Appl. Phys. B~{\bf 76} (2003) 143.

\bibitem{hansson-ktpabs} G. Hansson, H. Karlsson, S. Wang, F.
Laurell, Appl. Opt.~{\bf 39} (2000) 5058.
\bibitem{goudarzi} F.T.-Goudarzi, E. Riis, Opt. Commun.~{\bf 227}
(2003) 389.
\bibitem{zondy-ags} A. Douillet, J.-J. Zondy, A. Yelisseyev, S.
Lobanov, L. Isaenko, J. Opt. Soc. Am. B~{\bf 16} (1999) 1481.

\bibitem{bierlein} J.D. Bierlein, H. Vanherzeele, J. Opt. Soc. Am.
B~{\bf 6} (1989) 622.
\bibitem{B&K} G.D. Boyd, D.A. Kleinman, J. Appl. Phys.~{\bf 39}
(1968) 3597.
\bibitem{zondycomp} J.-J. Zondy, Opt. Commun.~{\bf 81} (1991) 427.

\bibitem{fradkin} K. Fradkin, A. Arie, A. Skliar, G.
Rosenman, Appl. Phys. Lett.~{\bf 74} (1999) 914.
\bibitem{kato2002} K. Kato, E. Tanaka, Appl. Opt.~{\bf 41} (2002)
5040.
\bibitem{ashkin} A. Ashkin, G.D. Boyd, J.M. Dziedzic, IEEE J.
Quantum Electron.~{\bf QE-2}(6) (1966) 109.

%\bibitem{polzik} E.S. Polzik, H.J. Kimble, Opt. Lett.~{\bf 16}
%(1991) 1400.
\bibitem{absolute} J.-J. Zondy, D. Touahri, O. Acef, J. Opt. Soc.
Am. B~{\bf 14} (1997) 2481.
\bibitem{chu-ktp} D.K.T. Chu, J.D. Bierlein, R.G. Hunsperger, IEEE
Trans. Ultrason. Ferroelec.\verb+&+ Freq. Control~{\bf 39} (1992)
683.

\bibitem{pelz} M. Pelz, U. B\"ader, A. Borsutzky, R. Wallenstein,
J. Hellstr\"om, H. Karlsson, V. Pasiskevicius, F. Laurell, Appl.
Phys. B~{\bf 73} (2001) 663.
\bibitem{boulanger} B. Boulanger, J.P. F{\`e}ve, G. Marnier, C.
Bonnin, P. Villeval, J.-J. Zondy, J. Opt. Soc. Am. B~{\bf 14}
(1997) 1380.
\bibitem{shoji} I. Shoji, T. Kondo, A. Masayuki, M. Shirane, R.
Ito, J. Opt. Soc. Am. B~{\bf 14} (1997) 2268.
\bibitem{eger} A. Arie, G. Rosenman, V. Mahal, A. Skliar, M. Oron,
M. Katz, D. Eger, Opt. Commun.~{142} (1997) 265.



\end{thebibliography}
\end{document}